\documentclass[pra,twocolumn,groupedaddress,shownopacs,amssymb]{revtex4-2}
\usepackage{graphicx,psfrag,amsmath,amssymb}
\usepackage[usenames, dvipsnames]{color}
\usepackage{braket}
\usepackage{amsmath}
\usepackage{amssymb}
\usepackage{float}
\usepackage{gensymb}
\usepackage{makeidx}
\usepackage{mathtools}

\usepackage[normalem]{ulem}
\usepackage[T1]{fontenc}
\usepackage{balance}
\usepackage{flushend}
\usepackage{lineno}
\usepackage{epsfig}
\usepackage{subfigure}
\usepackage{mathtools}
\usepackage{setspace}
\usepackage{bm}
\usepackage{ulem}
\usepackage{times}
\usepackage{enumitem}
\usepackage{xcolor}
\usepackage{multirow}
\usepackage{soul}
\usepackage{dcolumn}
\usepackage{balance}
\usepackage{epstopdf}
\usepackage{braket}
\usepackage{color}
\usepackage{diagbox}
\usepackage[utf8]{inputenc}
\usepackage{hyperref}
\hypersetup{
    colorlinks=true,
    linkcolor=Blue,
    citecolor=Blue,
    filecolor=Blue,
    urlcolor=Blue,
    }
\begin{document}
\title{Quasiparticle modes across soliton transition with density-dependent gauge field in optical lattices}
\author{Poorava Kumar Meena}
\affiliation{Department of Physics, Central University of Rajasthan, Ajmer - 305817, India}
\author{Kuldeep Suthar}
\affiliation{Department of Physics, Central University of Rajasthan, Ajmer - 305817, India}

\date{\today}
\begin{abstract}
 Tunneling of ultracold bosons confined in a one-dimensional optical lattice results in a nontrivial gauge field depending on density-difference between the sites involved. The density dependent tunneling (DDT) causes a first-order quantum phase transition from the Bose-Einstein condensate to the localized soliton. Here, we examine the low-lying quasiparticle mode evolution across the transition in the weakly-interacting limit. To this end, we employ the discrete Bogoliubov theory and dispersion curves to reveal an increase in the quasiparticle energies of dipole (and higher) excitations with DDT in the soliton phase while preserving the zero-energy mode. This is due to the decrease in effective tunneling, resulting in a larger energy cost to move the soliton, and causes faster dipole oscillations. The repulsive on-site atomic interaction further shifts the critical gauge field of DDT to a larger value by stabilizing the lattice soliton. The latter is corroborated by a phase diagram in the complex gauge field plane and mode energy gap of quasiparticles. We further show that the soliton does not exhibit Bloch oscillations as the effective momentum becomes density-dependent and deforms its internal structure. The dynamical response of the trap quench does not excite the width of localized wave-packet, and the breathing oscillations are suppressed. The latter two dynamical properties of the condensate uniquely contrast the soliton state due to the density-dependent gauge field.
\end{abstract}

\maketitle
\section{\label{intro}Introduction}
 Ultracold atoms in optical lattices  \cite{morsch_06,smerzi_03,rey_04,bloch_08,rapp_12} provide an ideal platform to simulate the low-energy behavior of condensed matter models~\cite{gross_17,schafer_20,zhao_20,lewenstein_07,lewenstein_12}. The trapping of atoms in discrete lattices provides a versatile tool to study many-body quantum and complex phenomena. The tuning of the lattice depth controls the competition of kinetic energy (hopping) and onsite repulsion, and results in the quantum phase transition from the superfluid (SF) phase to the Mott insulator (MI) phase \cite{zwerger_03,greiner_02,oosten_03,sengupta_05,dupuis_09,kashurnikov_02,xie_04,eckardt_05}. This setup is an excellent environment for investigating geometric frustration in low-dimensional systems that enables careful manipulation of the strength and sign of hopping rate. Laser-assisted hopping could result in a density-dependent gauge field \cite{barbiero_19,faugno_23,zhang_22,greschner_14}, in which the phase of hopping is dynamically dependent on the local particle occupation. In solid-state systems, the density-driven interaction processes can not be studied systematically due to the lack of direct control over the electronic density and interaction strengths. Moreover, the complexity of quantum materials \cite{hubbard_63} hinders the realization of interactions-induced phenomena such as bond-charge interaction \cite{hirsch_89}. The quantum gases are clean, defect free, and allow for direct observation of phenomena by controlling laser parameters and the Feshbach resonance \cite{chin_10,dickershied_05}. In the weakly-interacting limit of quantum gases in the optical lattice potential, the competition of particle occupancy with single-particle hopping can lead to instability and novel phase transition to localized solutions \cite{faugno_23}.       

Solitons are localized non-dispersive wave-packets that preserve their shape in propagation due to a balance between dispersion and non-linearity~ \cite{trombettoni_01,oxtoby_07,malomed_96,Papacharalampous_03,gonzalez_06,pankov_10,kevrekidis_01,becker_08}. The bright matter-wave solitons are localized density peaks on a low-density background, and have been the subject of extensive theoretical and experimental investigations. In lattice systems, discrete breathers are localized excitations that have been widely studied as solutions of non-linear systems~\cite{flach_08}. These lattice breathers in quasi-one-dimensional Bose-Einstein condensate (BEC) have been realized by quenching the attractive interaction strength~\cite{luo_20}. More recently, single-site and multi-site discrete bright soliton with attractive interaction have been observed in optical lattice by quenching the interaction strength and potential~\cite{cruickshank_25,cruickshank_26}. Various ways to excite the soliton mode have been proposed, such as spatial and time-dependent modulation, the phase-imprinting method \cite{burger_99}, and the quench of interactions~\cite{dicarli_19,khaykovich_02,strecker_02}. The fast periodic time variation of nonlinearity could completely suppress inter-site hopping and form compacton excitations \cite{dambroise_15,abdullaev_10}. Moreover, the nature of the interaction can be dynamically determined by the density gradient of the wavefunction, and thus such a density-dependent gauge field admits novel phenomena, including the coexistence of plane-wave (PW) condensate and the soliton state. The spontaneous symmetry breaking and consequently zero energy modes of a bright solitary wave in a one-dimensional toroidal trap have recently been studied~\cite{kavoulakis_26}. However, the quasiparticle excitations across the condensate to (the bright) soliton under a density-dependent gauge field have not yet been explored.    

In the present work, we examine the low-lying quasiparticle excitations and the evolution of their energies in the presence of a density-dependent gauge field. We use the discrete density difference dependent nonlinear Schr\"odinger equation (D$^{4}$NLSE) and Bogoliubov theory~\cite{bogoliubov_47,rey_03,fetter_72} to obtain the ground state solutions and quasiparticle mode energies. The system exhibits PW condensate to a localized bright soliton transition with field strength. We further introduce onsite repulsion to reveal the shift in the critical values towards higher value of gauge field. This is reflected in the quasiparticle excitation spectrum as hardening of the non-zero excited mode and the emergence of the quasiparticle mode energy gap. We show the momentum space correlation that destroys in time dynamics when the imaginary gauge coupling exceeds a threshold value predicted by dispersion relation, which signals the onset of modulational instability. The nonlinear gauge term couples the center-of-mass motion of the soliton to its internal shape, and thus the soliton does not exhibit Bloch oscillations \cite{kolovsky_09, morsch_01,salerno_08} in contrast to the dynamical response of the condensate. The quench of the trap strength causes breathing oscillations \cite{carretero_02, chen_22} in the condensate phase, whereas no sustained breathing oscillations develop in the localized soliton state. 

The remainder of the paper is organized as follows. In Sec.~\ref{THEORY} we discuss the model Hamiltonian, D$^{4}$NLSE, Bogoliubov theory, and the single-particle dispersion relation. In Sec.~\ref{results} we present the quasiparticle excitations, dispersion curves, ground-state phase diagrams, and dynamical properties of the model Hamiltonian. Finally, we conclude in Sec.~\ref{conclusion}.

\section{Model Hamiltonian and Method}\label{THEORY}
Consider a Bose-Einstein condensate confined in a quasi-one-dimensional optical lattice and a background external harmonic trap. The radial trapping frequency of the external harmonic potential is much larger than that of the axial direction, such that the excitations along the tight (high frequency) axes are of high energies and the system remains in the ground state in the transverse direction. In the tight-binding limit, the Bose-Hubbard model Hamiltonian describing the system is       
\begin{eqnarray}
  \hat{H} &=&\sum_j \hat{b}^{\dagger}_{j+1}\Big[-J+\gamma(\hat{n}_{j+1}-\hat{n}_j)\Big] \hat{b}_j \nonumber \\
          &+&\sum_j\hat{b}^{\dagger}_{j}\Big[-J+\gamma(\hat{n}_{j}-\hat{n}_{j-1})\Big]\hat{b}_{j-1} + {\rm H.c.} \nonumber \\
          &+&\sum_j\left[(\epsilon_j - \mu)\hat{n_j} +\frac{U}{2}\hat{n}_{j}\left(\hat{n}_{j}-1\right)\right],
\label{eq:hami}
\end{eqnarray}
where $\hat{b}^{\dagger}_{j}$ ($\hat{b}_j$) is the bosonic creation (annihilation) operator of bosonic atoms at the $j$th lattice site, $J$ describes density-independent nearest-neighbour single-particle hopping, $\gamma$ describes density-difference-dependent hopping which can in general be a complex number, $\hat{n}_j=\hat{b}^{\dagger}_{j}\hat{b}_j$ is the number operator at the $j$th site. Here, $\epsilon_j=j^2\Omega$ is the external harmonic trapping potential with $\Omega = m\omega^{2}_{z}a^{2}/2$ as the strength of the potential depending on the trapping frequency and the atomic mass of the species. Here, $a$ is the lattice spacing. $U=2\sqrt{\lambda\kappa}\hbar^2Na_s/m$ \cite{suthar_15} is the strength of onsite interaction with $N$ is the total number of atoms in the lattice, $m$ is the mass of the atomic species, $\lambda=\omega_x/\omega_z$ and $\kappa=\omega_y/\omega_z$ are anisotropy parameters along the $y$  and $z$ directions, and $a_s$ is the s-wave scattering length which is repulsive in the present work. $\mu$ is the chemical potential of the system that controls the number of atoms. Although $U$ is repulsive, the sign of the hopping-driven interaction could be repulsive or attractive and is not determined a priori but depends on the density-difference of lattice occupancy during the hopping process.  

The equation of motion of the model Hamiltonian can be obtained using the Poisson bracket relation. 
To study the quantum mechanical weakly-interacting system in the mean-field limit, the Poisson bracket is replaced with a commutator bracket. The resulting discrete density-difference-dependent nonlinear Schr\"odinger equation is
\begin{eqnarray}
 i\hbar\frac{d\Psi_j}{dt} &=& \{-J + \gamma(2n_j - n_{j-1})\}\Psi_{j-1} - \gamma\Psi_j^2 \Psi_{j+1}^* \nonumber \\
                     &+& \{-J - \gamma^*(2n_j - n_{j+1})\}\Psi_{j+1} + \gamma^*\Psi_j^2 \Psi_{j-1}^* \nonumber \\  
                     &+& (\epsilon_j-\mu) \Psi_j + U|\Psi_j|^2\Psi_j,
\label{d4nlse}
\end{eqnarray}
where $\Psi_{j}$ is a complex amplitude corresponding to the Hubbard annihilation operator and $\Psi_{j}^{*}$ corresponds to the creation operator at the $j$th site. The stationary ground state 
of the system is obtained using imaginary-time propagation with the state evolving as $\Psi_j(t) = e^{-i\mu t}\Psi_j(0)$. Note that the usual DNLSE includes the onsite repulsion, while here the nearest-neighbour hopping of the atoms also contributes to the effective interaction. 

To examine the quasiparticle excitation spectra of the system, we employ the mean-field approximation~\cite{griffin_96} to decompose the Bose field operator or the corresponding complex amplitude at each site in terms of the mean-field $c_{j}$ representing the macroscopic condensate and fluctuation operator $\tilde{\varphi}_{j}$, as $\Psi_{j}=c_{j}+\tilde{\varphi}_{j}$, where the second term averages to zero such that $\langle\tilde{\varphi}_{j}\rangle=0$. Similarly, the respective creation operator is defined at a given site. The chemical potential of the system can be 
obtained following the normalization of complex amplitudes and is given by 
\begin{eqnarray}
  \mu &=& \left[-J+\gamma(2|c_j|^2-|c_{j-1}|^2\right]c_{j-1}c_j^*-\gamma |c_j|^2c_jc_{j+1}^* \nonumber \\
        &&\left[-J-\gamma^*(2|c_j|^2 -|c_{j+1}|^2\right]c_{j+1}c_j^*+\gamma^*|c_j|^2c_jc_{j-1}^* \nonumber \\
        &&+\epsilon_{j}|c_j|^2 + U|c_j|^4.
\end{eqnarray}
The equation of motion of the fluctuation operator representing the non-condensate part is $i\hbar\dfrac{d\tilde{\varphi}_{j}}{dt} = i\hbar\dfrac{d}{dt}(\Psi_{j} - c_{j})$. We further linearize the equation of fluctuation operator by retaining the terms upto first order in the fluctuation operator. The higher-order terms represent the role of thermal fluctuations at finite temperatures and are not included in the present work.  The expectation of quadratic fluctuation operators $\langle\tilde{\varphi}_j^\dagger\tilde{\varphi}_j\rangle = \tilde{n}_{j}$ is the local non-condensate density, and $\langle\tilde{\varphi}_j\tilde{\varphi}_j\rangle = \tilde{m}_{j}$ is the anomalous density. The latter is the expectation value of an unequal number of creation and annihilation operators and does not play a significant role in the excitations. In order to obtain the gapless excitation spectrum for the repulsive condensate, the anomalous term is omitted. At zero temperature, thermal fluctuations are absent and thus the non-condensate density $\tilde{n}$ is omitted. The linearized equation of motion of the fluctuation operators
\begin{align}
i\hbar\frac{d\tilde{\varphi}_{j}}{dt} &= -J\{\tilde{\varphi}_{j+1}+\tilde{\varphi}_{j-1}\}+ 2\gamma\{|c_{j}|^{2}\tilde{\varphi}_{j-1} +c_{j}^{*}c_{j-1}\tilde{\varphi}_{j} \nonumber \\
&+c_{j}c_{j-1}\tilde{\varphi}_{j}^{\dagger}{}  -\gamma\{2|c_{j-1}|^{2}\tilde{\varphi}_{j-1}+c_{j-1}^{2}\tilde{\varphi}_{j-1}^{\dagger}{} +c_{j}^{2}\tilde{\varphi}_{j+1}^{\dagger} \nonumber \\
&+2c_{j}c_{j+1}^{*}\tilde{\varphi}_{j}\} -2\gamma^{*}\{|c_{j}|^{2}\tilde{\varphi}_{j+1} +c_{j}^{*}c_{j+1}\tilde{\varphi}_{j}+c_{j}c_{j+1}\tilde{\varphi}_{j}^{\dagger} \nonumber \\
&+\gamma^{*}\{2|c_{j+1}|^{2}\tilde{\varphi}_{j+1}+c_{j+1}^{2}\tilde{\varphi}_{j+1}^{\dagger} +c_{j}^{2}\tilde{\varphi}_{j-1}^{\dagger}+2c_{j}c_{j-1}^{*}\tilde{\varphi}_{j}\}  \nonumber \\ &+(\epsilon_j-\mu)\tilde{\varphi}_j + U(2|c_j|^2\tilde{\varphi}_j + c_j^2\tilde{\varphi}^\dagger_j),
\end{align}
which stems from the leading (second-order in fluctuation operator) quadratic Hamiltonian as the contributions from first-order vanish. To diagonalize the quadratic Hamiltonian, we use the Bogoliubov transformation to define the fluctuation operator at the $j$th site is   
\begin{equation}
  \tilde{\varphi}_{j} = \sum_{l}[u_{j}^{l}\hat{\alpha}_{l}e^{-iE_{l}t/\hbar}-v_{j}^{l*}\hat{\alpha}_{l}^{\dagger}e^{iE_{l}t/\hbar}],
 \label{bogo_trans}
\end{equation}
where $u_j^l$ and $v_j^l$ are the quasiparticle and quasihole amplitudes or mode functions for the $l$th quasiparticle mode with mode energy $E_l$. $\hat{\alpha}_l$ and $\hat{\alpha}_l^\dagger$ are the quasiparticle annihilation and creation operators. The quasiparticle and quasihole mode amplitudes follow the following normalization conditions
\begin{subequations}
\begin{eqnarray}
  \sum_{j} \left( u_{j}^{*l} u_{j}^{l'} - v_{j}^{*l} v_{j}^{l'} \right) &=& \delta_{ll'}, \\
  \sum_{j} \left( u_{j}^{l} v_{j}^{l'} - v_{j}^{*l} u_{j}^{*l'} \right) &=& 0.
\end{eqnarray}
\label{mode_norm}
\end{subequations}
Applying the Bogoliubov transformation [Eq.~\eqref{bogo_trans}] for the fluctuation operator with the above conditions and collecting the prefactors of $e^{-iE_{l}t/\hbar}$ and $e^{iE_{l}t/\hbar}$, we obtain the coupled Bogoliubov-de Gennes (BdG) equations, which in matrix eigenvalue form can be written as
\begin{equation}
  \begin{pmatrix} 
    \mathcal{L} & \mathcal{M} \\ -\mathcal{M}^{*} & -\mathcal{L}^{*} \end{pmatrix} \begin{pmatrix} u^{l}_{j} \\  v^{l}_{j} \end{pmatrix} = E_{l} \begin{pmatrix} u^{l}_{j} \\ v^{l}_{j}
  \end{pmatrix}.
\end{equation}
The matrix elements of the above matrix are given by
\begin{eqnarray}
  \mathcal{L}_{j,j} &=& 2\gamma c_{j}^*c_{j-1}-2\gamma^{*}c_{j}^{*}c_{j+1}+2\gamma^{*}c_{j}c_{j-1}^{*} \nonumber \\ 
                    &&-2\gamma c_{j}c_{j+1}^{*} + \epsilon_j-\mu + 2U|c_j|^2\nonumber\\
  \mathcal{L}_{j,j-1} &=& -J+2\gamma|c_{j}|^{2}-2\gamma|c_{j-1}|^{2} \nonumber\\
  \mathcal{L}_{j,j+1} &=& -J-2\gamma^{*}|c_{j}|^{2}+2\gamma^{*}|c_{j+1}|^{2} \nonumber\\
  \mathcal{M}_{j,j} &=&\hspace{0.5em}2\gamma^{*}c_{j}c_{j+1} - 2\gamma c_{j}c_{j-1} - Uc_j^2  \nonumber\\
  \mathcal{M}_{j,j-1} &=& \hspace{0.2em}\gamma c_{j-1}^{2}-\gamma^{*}c_{j}^{2} \nonumber\\
  \mathcal{M}_{j,j+1} &=& \hspace{0.2em} \gamma c_{j}^{2}-\gamma^{*}c_{j+1}^{2}.
  \label{eq:bdg}
\end{eqnarray}

Unlike single-particle hopping, the complex density-dependent hopping terms appear in the off-diagonal blocks as well, because it couples with the particle occupancy and contributes to the effective interaction. The above coupled BdG equations are self-consistently solved to obtain the lattice distribution of quasiparticle amplitudes and corresponding quasiparticle energies. 

To analyze the stability of the PW condensate in the presence of a density-difference gauge field, the dispersion relation \cite{faugno_23} can be obtained by introducing a small mean-field fluctuation around the zero-momentum condensate.   
\begin{equation}
  \Psi_{j}(t) = e^{-i\mu t} \left( \frac{1}{\sqrt{L}} + \delta_j(t) \right),
\end{equation}
where $L$ is the total number of lattice sites, $|\delta_{j}| \ll 1/\sqrt{L}$ is a small fluctuation to a stationary solution $\mu=-2J$. The local density at $j$th site $n_j\approx\frac{1}{L} + \frac{1}{\sqrt{L}}(\delta_j + \delta_j^{*})$, where higher order terms of $\delta$ are neglected. We use the above approximation in D$^4$NLSE with $U=0$ to obtain the equation of motion for the fluctuation which reads as
\begin{equation}
i\hbar \frac{d\delta_j}{dt} = -J(\delta_{j+1} + \delta_{j-1} - 2\delta_j) + \frac{2i\gamma_I}{L}(2\delta_j^* - \delta_{j-1}^* - \delta_{j+1}^*).
\label{fluc_k}
\end{equation}
The Bogoliubov transformation in the momentum space is $\delta_j(t) = u_{j} e^{i(kj - \omega t)} + v^{*}_{j}e^{-i(kj - \omega t)}$, with $\delta_{j\pm1}$ possesses the $e^{\pm ik}$ phase for the quasiparticle, while the quasihole acquires the opposite phase. Substitution of Fourier transformed fluctuation in the above Eq.~\eqref{fluc_k} leads to coupled eigenvalue equations of $u$ and $v$, where the Bogoliubov Hamiltonian matrix is 
\begin{equation}
  \mathcal{H}_{\rm BdG} = 2J(1 - \cos ka)\mathbb{1} - \frac{4\gamma_{I}}{L} \sin ka~~\sigma_{y}, 
\end{equation}
where $\mathbb{1}$ is an identity matrix and $\sigma_{y}$ is the Pauli spin matrix. It should be noted that the imaginary part $\gamma_{I}$ retained, whereas the contribution of the real part $\gamma_{R}$ cancels out. The diagonalization of the above BdG matrix results in excitation energy in the momentum space
\begin{equation}
  \epsilon(k) = 2J\sqrt{1 - \frac{4\gamma_I^2}{J^2L^2}} (1 - \cos ka).
  \label{eq.dispersion}
\end{equation}
The characteristic features of the above dispersion relation are as follows: The dispersion is quadratic at a smaller momentum. The gauge field strongly affects the dispersion through the imaginary component. Above a  critical value $\gamma_{I} > JL/2$ the dispersion becomes imaginary (complex), which is the signature of modulational instability. The transition to unstable condensate is system-size dependent. Thus, the stability of the quantum phases of the model Hamiltonian depends on $\gamma_{I}$ while the transition between the phases is determined by both components of $\gamma$.    
\section{Results and Discussions}\label{results}
\subsection{Numerical details}
We first scale the D$^4$NLSE by the recoil energy $E_{R} = \hbar^2\pi^2/2ma^2$ which serves as a characteristic energy scale with $m$ being the mass of the atomic species. The lattice spacing (constant) $a$ is the length scale. The scaled equation is solved using the fourth-order Runge-Kutta method to obtain the equilibrium ground state. The initial complex amplitude at the $j$th lattice site is chosen as a Gaussian amplitude profile with a random phase, which is given as $c_{j} = [e^{-j^2/2\sigma^2}/(\pi\sigma^2)^{1/4}] \times e^{2\pi i \theta_j}$ with $\sigma$ as the width of the Gaussian profile, and the value of $\theta_{j}$ is a random phase. We further use imaginary-time propagation~\cite{steel_98} to solve time-independent D$^4$NLSE to obtain the stationary solution. The number of iterations is considered such that the solution converges to the complex amplitude with a tolerance of the order of $10^{-6}$. The ground state density and the corresponding chemical potential are further used in computing the matrix elements of the BdG matrix. In general, matrix elements can be a complex number, resulting in complex eigenvalues. The BdG matrix has the dimension of $2L \times 2L$ and could be non-Hermitian and non-symmetric. To examine the excitation spectrum, we diagonalize the matrix and obtain the quasiparticle (quasihole) amplitudes $u_j^l (v_j^l)$ as eigenstates and quasiparticle energies as eigenenergies. We employ the {\tt ZGEEV} subroutine from the LAPACK library~\cite{anderson_99}.

\subsection{Quasiparticle excitations}
 To examine the quasiparticle excitations of the model Hamiltonian, we first study the ground state density profiles and analyze the variation of ground state energy and chemical potential. As density-dependent hopping is a complex value, we further examine the evolution of mode energies as a function of real and imaginary parts of $\gamma$. 
\begin{figure}[h]
  \includegraphics[width=\linewidth]{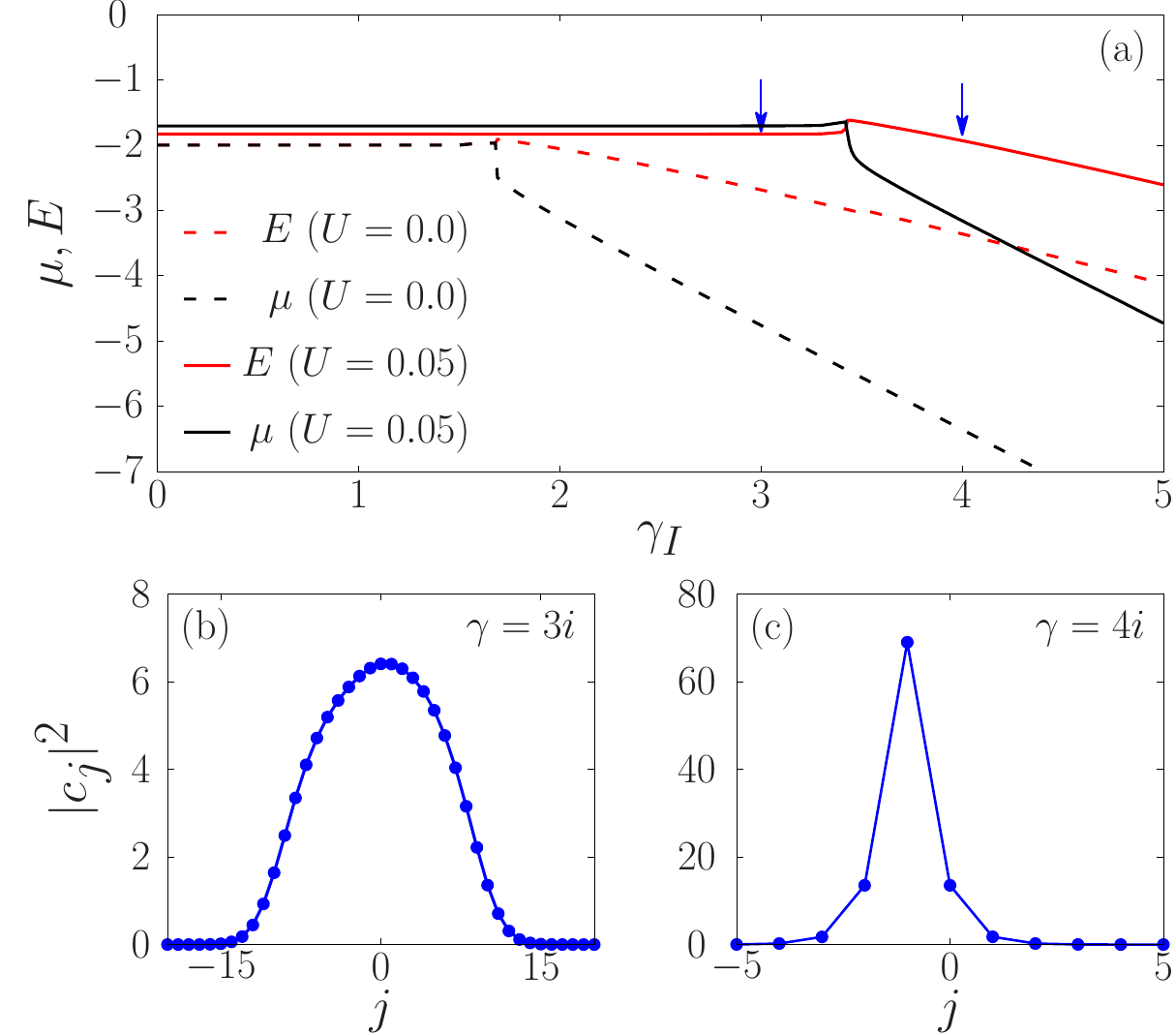}
  \caption{(a) The ground state energy $E$ and chemical potential $\mu$ obtained from the imaginary time propagation as a function of imaginary part of gauge coupling $\gamma_{I}$. Here the real part is 
  $\gamma_{R}=0$. The energy and $\mu$ for the non-interacting uniform case $U=0$ are shown with dashed lines while those of interacting case with $U=0.05$ are shown with solid lines. The $E$ and $\mu$ remains steady in PW (Bose-Einstein) condensate phase but begins to decrease at a critical $\gamma_{I}$ as the system enters into localized soliton state. The critical $\gamma_{I}$ shifts to higher values (compared to $U=0$) for interacting systems. The ground state density profiles at two representative $\gamma_{I}$ values are shown in (b) corresponding to BEC phase and (c) soliton state for $U=0.05$. The arrows in (a) indicate the respective $\gamma_{I}$ values.}
  \label{ener_mu}
\end{figure}

 \subsubsection{Ground state energy and chemical potential}
 We first obtain the stationary state for the quasi-1D Bose gas trapped in a lattice potential. The number of lattice sites is taken as $L=110$. The single-particle nearest-neighbour hopping is set as $J=1$ in the present work. 

To examine the ground states of the Hamiltonian [Eq.~\eqref{eq:hami}], we analyze the variation in the ground state energy and chemical potential as a function of purely imaginary $\gamma$, i.e. $\gamma_{I}$ while keeping $\gamma_{R}$ to zero. The $E$ and $\mu$ of a non-interacting uniform case are shown by dashed lines in Fig.~\ref{ener_mu} which shows that these remain constant with $\gamma_{I}$ in the plane wave condensate phase but decrease as the gauge coupling ramps above a critical value~\cite{faugno_23}. The steady value in the PW condensate is $-2$, following the dispersion relation as $J=1$. The critical $\gamma_{I}$ is $1.67$ above which the soliton is a ground state. The variation in energy and $\mu$ suggest that it is a first-order transition. Thus, the gauge coupling leads to a condensate-to-soliton phase transition, as evident in the change in $E$ and $\mu$. This can be understood as follows: The ground state with $k=0$ is a homogeneous plane-wave condensate with $\mu=E\approx-2J$. The gauge coupling driven localized solution can be obtained by using a three-site ansatz 
\begin{equation}
  {\Psi}_j= \sqrt{\alpha},~~{\Psi}_{j\pm1}=\sqrt{(1-\alpha)/2}e^{i\theta_{\pm}},
\end{equation}
where $\alpha$ and $\theta_{\pm}$ are variational parameters. The prefactors of the amplitude ensure the normalization condition. The variational parameters can be obtained by minimizing the Hamiltonian. This leads to the energy functional as
\begin{equation}
  \begin{split}
  {\mathcal{E}} ={} & 2\sqrt{\alpha}\sqrt{\frac{1-\alpha}{2}} \left[ \left(-J + \gamma\frac{3\alpha - 1}{2}\right)\cos(\theta_-) \right. + \\
  & \left. \left(-J - \gamma\frac{3\alpha - 1}{2}\right)\cos(\theta_+) \right].
  \end{split}
  \label{eq:energy_functional}
\end{equation}
The variational parameters are obtained as $\alpha=0.81$ and $\theta_{\pm}=m\pi$ with $m=0,\pm1,\pm2,..$. Among the four possible combinations of $\theta_{+}$ and $\theta_{-}$, the lowest possible value of $\mathcal{E}$ determines the ground state, which is satisfied for $\theta_+= 2m\pi, \theta_-= (2m+1)\pi$. The substitution of variational parameters in an energy functional results in  the energy of the soliton state as $E_{\rm{sol}}=-0.792\gamma$. Comparing $E_{\rm{sol}}$ with the ground state energy of the plane wave condensate $E_{\rm{pwc}}=-2J$ gives the desired critical value of $\gamma$. This value is $\gamma_{c}=2.53$~\cite{faugno_23}. The prediction of a higher value in the analysis is justified because the soliton could be localized in more than three lattice sites.  

The presence of repulsive onsite interaction enhances the effective interaction. Here, we consider the harmonic confinement with strength $\Omega=0.001$. This causes a shift in the critical gauge coupling towards a higher value. This is evident from the variation of $E$ and $\mu$ at $U=0.05$ (in units of $E_{R}$). Although the qualitative behaviour remains the same, the slight upward shift in $\mu$ and $E$ for BEC is attributed to atomic interaction and confinement. The critical value of $\gamma_I$ shifts to $3.45$ for the parameters considered. Hence, the gauge coupling driven soliton transition is present even with the weak many-body interaction. In the weakly-interacting limit, the interaction can be treated as a small perturbation that leads to a first-order shift in the critical coupling value to $\gamma_{c} = 4.65$ at $U=0.05$. Thus, the finite value with $U$ confirms the numerical observation of the interacting phase transition. At lower $\gamma_{I}$ the system exhibits Bose-Einstein condensate, and above the critical value of gauge coupling, the system makes a transition to the localized soliton state. The density profiles of the ground state for these two phases are presented for $\gamma_{I} = 3$ and $\gamma_{I} = 4$ in Fig.~\ref{ener_mu} (b) and Fig.~\ref{ener_mu} (c), respectively.  

\subsubsection{Quasiparticle mode evolution}
\begin{figure}[h]
  \includegraphics[width=\linewidth]{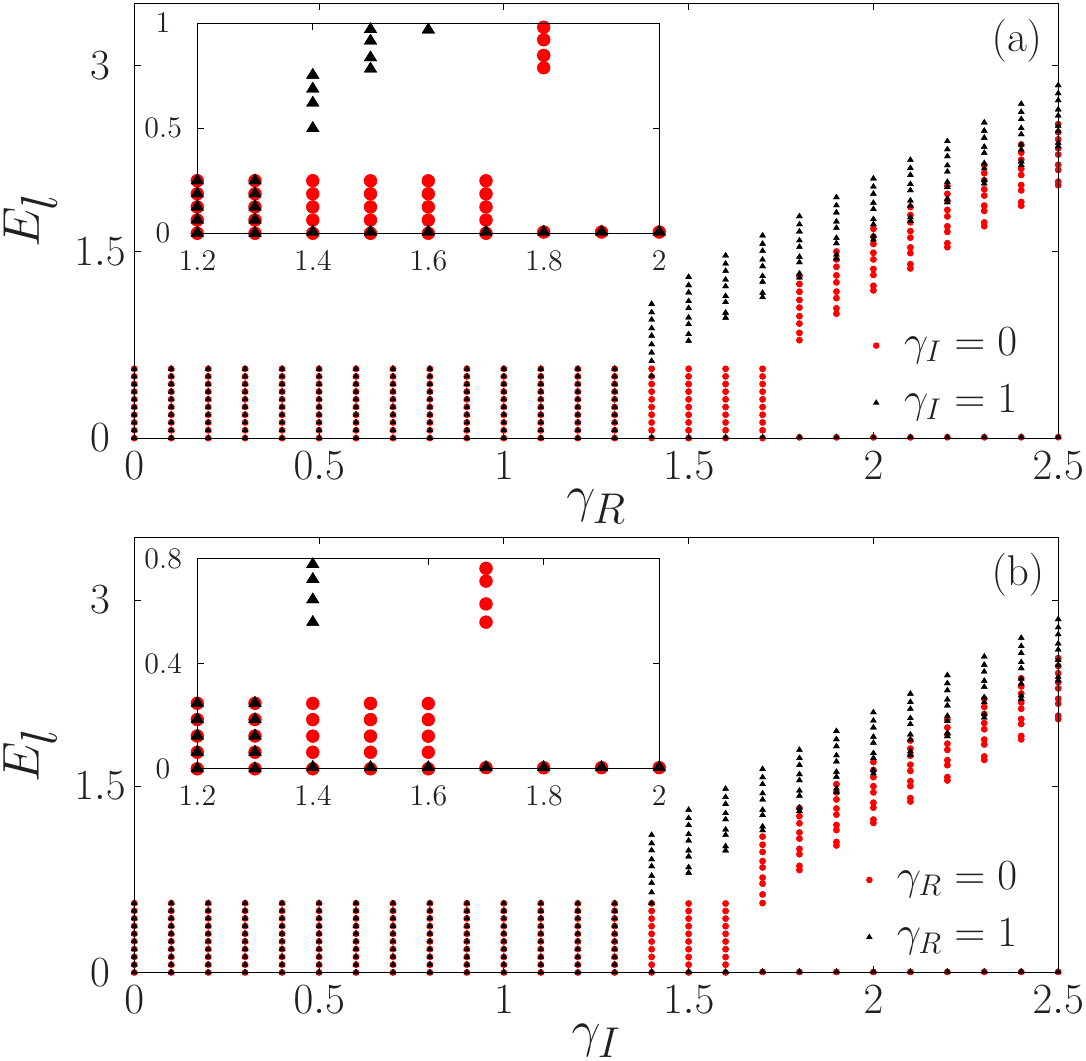}
  \caption{The evolution of quasiparticle mode energies for the non-interacting case. The upper panel (a) shows the mode energy evolution with $\gamma_{R}$ for fixed imaginary parts $\gamma_{I} 
           = 0$ and $\gamma_{I} = 1$. The lower panel shows the mode energies with $\gamma_{I}$ for fixed real parts $\gamma_{R} = 0$ and $\gamma_{R} = 1$. The insets show the magnifying evolution of mode energies near the phase transition. At the localized phase transition, the low-lying modes get hardened and gap increases with $\gamma$. The shift in the critical value with finite $\gamma_{R}$ ($\gamma_{I}$) indicates the $\gamma$-dependent phase transition. The mode energies and gauge coupling are in units of the recoil energy $E_{R}$. Here, $J=1$ and periodic boundary condition is assumed.}
  \label{fig:mode_u=0}
\end{figure}

We now turn to investigate the low-lying quasiparticle modes as a function of gauge coupling. The evolution of the mode energies with real and imaginary $\gamma$ are shown in Fig.~\ref{fig:mode_u=0}(a) and Fig.~\ref{fig:mode_u=0}(b), respectively. 
 
We first examine the evolution of the energies of low-lying quasiparticle modes with real part of $\gamma$ at fixed $\gamma_{I} = 0$. The evolution of modes shows a distinct hardening of modes at a critical value of $\gamma_{R}$. In the plane-wave condensate of the single-particle system, the mode energies remain steady at lower values of $\gamma_{R}$. In the PW condensate phase, the system possesses one zero-energy and other finite-energy modes. At a critical value $\gamma_{R} = 1.78$, the low-lying quasiparticle modes are hardened. In the bright soliton phase, the localized wave-packet modifies the hopping contributions, and the deformation in such a localized mode costs significant energy of dipole oscillations. Unlike in continuum, where the soliton can move freely, the discrete system of lattice pins the soliton to a few sites near the center, which causes a finite energy cost in the excitation spectrum. As the strength of $\gamma_{R}$ increases, the degree of localization of the soliton is improved, thereby leading to a reduction of its width and a larger density gradient of the ground state. Near the transition, the hardening is evident by the evolution shown in the inset of Fig.~\ref{fig:mode_u=0}(a). We further find that at a finite fixed value of $\gamma_{I}$ ($\gamma_{I}=1$), the qualitative behaviour remains the same; however, the critical $\gamma_{R}$ of mode hardening shifts to lower values [cf. circles (red) and triangles (black) points]. The critical value is obtained as $\gamma_{R} = 1.38$. This is consistent with the critical $\gamma$ of the quantum phase transition from PW condensate to a localized soliton. 
 
In order to examine the dependence of the complex $\gamma$, we further study the evolution of the quasiparticle energies as a function of $\gamma_{I}$ at fixed $\gamma_{R}=0$ (purely imaginary). The mode evolution exhibits a similar trend of mode energies, i.e. steady $E_{l}$ in the condensate phase while increasing mode energies with $\gamma_{I}$ in the soliton phase. However, the critical $\gamma_{I}$ of mode hardening corresponding to the soliton transition is modified. In particular, the critical $\gamma_{I}$ is $1.69$, consistent with the variation of $E$ and $\mu$, presented in Fig.~\ref{ener_mu}. Thus, purely imaginary $\gamma$ affects the phase of effective hopping and alters the transition space. At finite $\gamma_{R}$ ($\gamma_{R}=1$), with complex $\gamma$ the mode hardening occurs at $\gamma_{I}=1.36$. Thus, with complex $\gamma$ both the real and imaginary parts contribute to the effective hopping, which results in an energetically favourable soliton state at a smaller critical $\gamma$ value. 
\begin{figure}[h]
    \includegraphics[width=\linewidth]{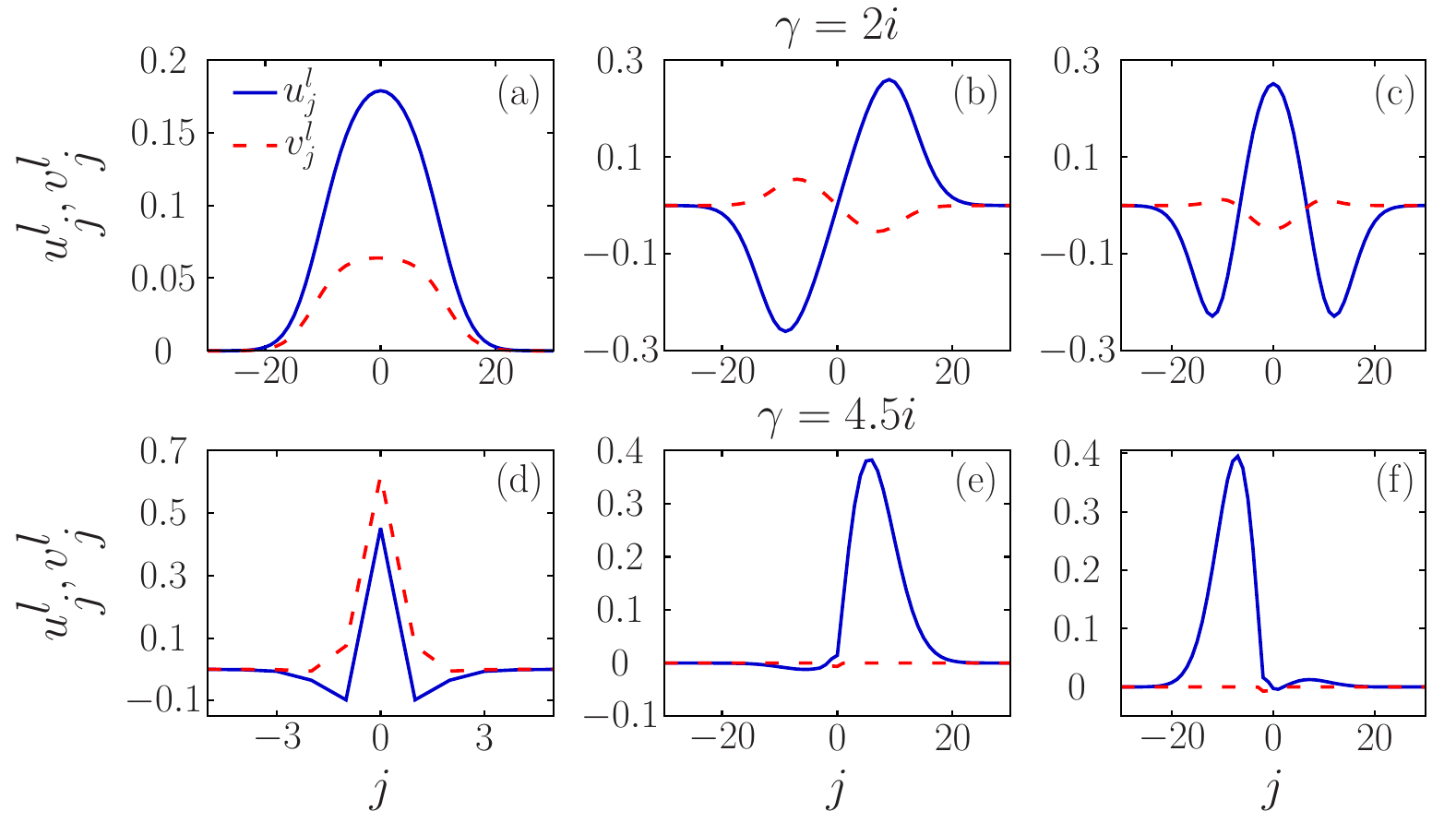}
    \caption{The lattice distribution of quasiparticle mode amplitudes of first three modes of D$^4$NLSE model with onsite repulsion. The modes with mode index $l=1$, $l=2$, $l=3$ are 
             characterized as zero-energy Goldstone mode, dipole mode, and breathing mode. The upper panel shows the modes for BEC phase with $\gamma = 2i$ while the lower panel shows the modes for soliton phase with $\gamma=4.5i$. The mode amplitudes are normalized as per Eq.~\eqref{mode_norm}.}
    \label{fig:amplitudes}
\end{figure}

Like in the previous subsection, we further examine the quasiparticle mode evolution in the presence of repulsive onsite interaction. The quasiparticle amplitudes of the first three low-lying modes in the two regime of $\gamma$ are shown in Fig.~\ref{fig:amplitudes}. In the BEC phase, these modes correspond to the zero-energy mode of spontaneous symmetry breaking, the dipole mode of center-of-mass oscillations, and breathing mode oscillations, respectively. The modes of the BEC phase are presented for $\gamma=2i$ and $U=0.02$. The zero-energy mode reflects the density in the ground-state, and the respective quasiparticle and quasihole amplitudes are illustrated in Fig.~\ref{fig:amplitudes}(a). The other two modes in Fig.~\ref{fig:amplitudes}(b) and Fig.~\ref{fig:amplitudes}(c) are dipole and breathing modes, respectively. For the larger $\gamma$, as the system transitions into the soliton state, the mode amplitudes are localized. The zero-energy mode is strongly localized near the center within a few sites and reflects the ground state [Fig.~\ref{fig:amplitudes}(d)]. At strong $\gamma$, the dipole mode lobes become unequal; in particular, the mode is prominently localized on one side of the trap center [Fig.~\ref{fig:amplitudes}(e)]. This strong asymmetry arises due to the coupling of amplitude and phase fluctuations in the presence of a Peierls phase of the gauge field $\gamma$\cite{gorg_19,greschner_15,keilmann_11}. This is also attributed to the slightly off-centered localized soliton, which breaks inversion symmetry [see Fig.~\ref{ener_mu}(c)]. The dipole mode is proportional to the discrete derivative ($c_{j+1} - c_{j-1}$), and the mode amplitude of the dipole excitation is determined by the slope of the soliton on each side. The derivative is larger on the side for which the soliton state is steeper, and accordingly the mode becomes more localized on one side. Similarly, the second nonzero mode also gets localized, as shown in Fig.~\ref{fig:amplitudes}(f).
\begin{figure}[h]
    \includegraphics[width=\linewidth]{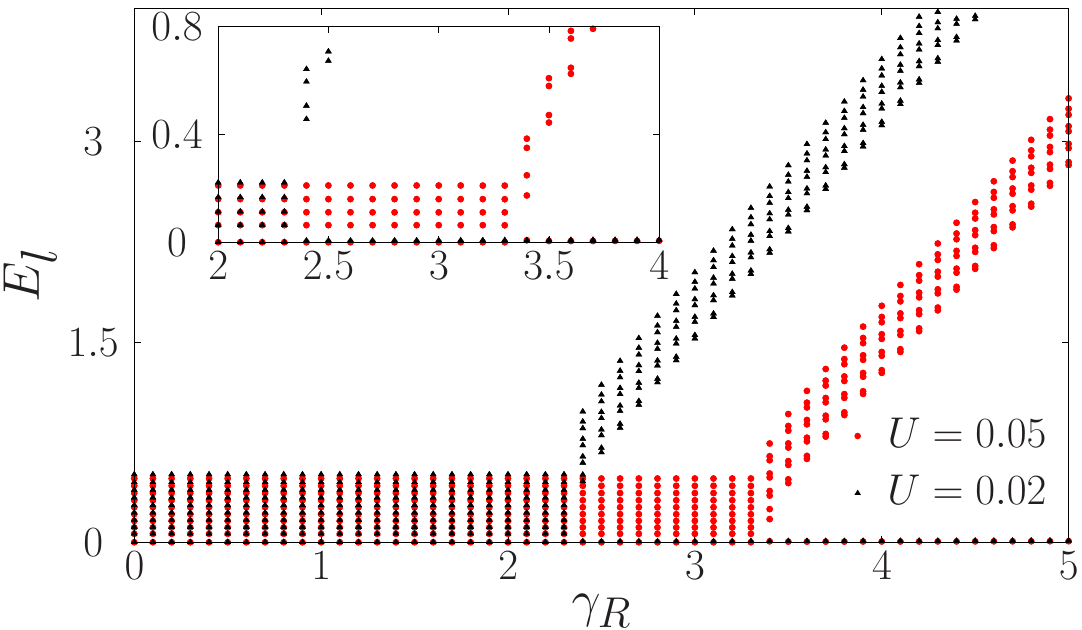}
    \caption{The evolution of quasiparticle mode energies for the interacting case. The figure shows the mode energy evolution with $\gamma_R$ for fixed imaginary part $\gamma_I=0$. The transition point increases with the inclusion of onsite interaction to $\gamma_R=2.34(3.39)$ for $U=0.02(0.05)$. The inset shows the magnifying evolution of mode energies near the phase transition. Here, the number of atoms $N=100$.}
    \label{fig:mode_U}
\end{figure}

We now examine the role of onsite repulsion in the phase transition between the BEC and the soliton state. In the weakly-interacting limit we consider the parameters $N=100$, and $U=0.02$ and $U=0.05$.  The evolution of the quasiparticle modes of the interacting case for two values of $U$ are shown in Fig.~\ref{fig:mode_U}. For brevity, we show the energies of the modes as a function of $\gamma_{R}$ with the imaginary part of $\gamma$ fixed to zero (purely real). We find that the quasiparticle spectra of the single-particle system is inherited with $U$. The qualitative behaviour of mode energies remains the same; however, the critical values shift due to the finite repulsion energy cost. For $U=0.02$, the hardening of modes occurs at $\gamma_{R}=2.34$, which is higher compared to the corresponding single-particle system. The higher critical $\gamma_{R}$ with finite $U$ is consistent with the analytical results of the three-site ansatz. We further find that the increase in the onsite interaction at $U=0.05$ shifts the soliton transition to a higher value at $\gamma_{R}=3.39$. Note that the onsite interaction favours spreading; thus it competes with the localization driven by gauge field. Therefore, the gauge field energy should be balanced by both kinetic (single-particle hopping) as well as onsite interaction. This results in a stronger $\gamma$ for the BEC phase before the system transitions into the localized soliton state. Furthermore, the critical field increases linearly with $U$ in the weakly-interacting limit. At a finite $\gamma_{I}$, the qualitative behaviour of the soliton transition is similar to the non-interacting case, hence not shown here. Hence, the gap of the quasiparticle spectrum serves as an ideal indicator of localization transition, which we discuss in the following subsection to map out the parameter space of the phases.
 \begin{figure}[h]
    \includegraphics[width=\linewidth]{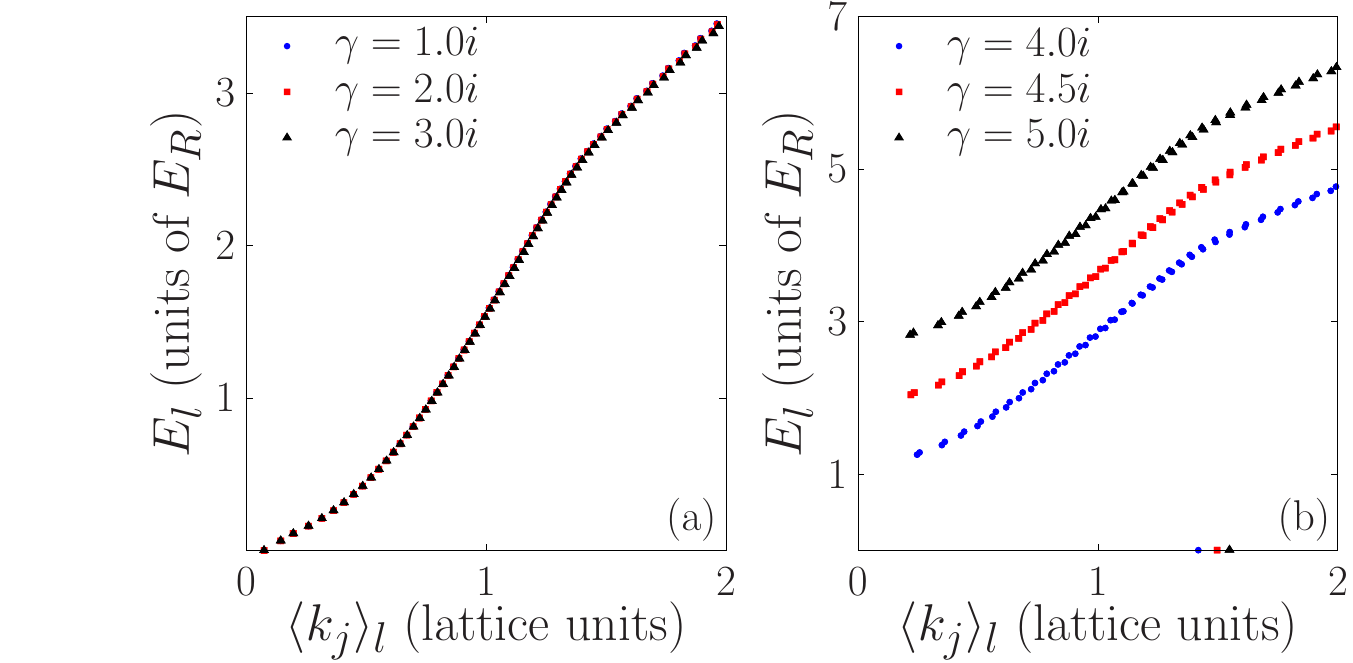}
    \caption{The discrete BdG quasiparticle dispersion curves corresponding to (a) BEC phase at $\gamma = 1.0i, 2.0i,$ and $3.0i$. The smooth, monotonic dispersion curves closely resemble the 
             standard gapless Bogoliubov spectrum of a stable lattice condensate. (b) Quasiparticle dispersion curves in the highly localized bright soliton phase for larger imaginary gauge couplings $\gamma = 4.0i, 4.5i$ and $5.0i$. The curve shows a significant energy gap of excited modes due to the reduction of effective inter-site tunneling and pinning effects.}
    \label{dispersion_curves}
\end{figure}
\begin{figure}[h]
    \includegraphics[width=\linewidth]{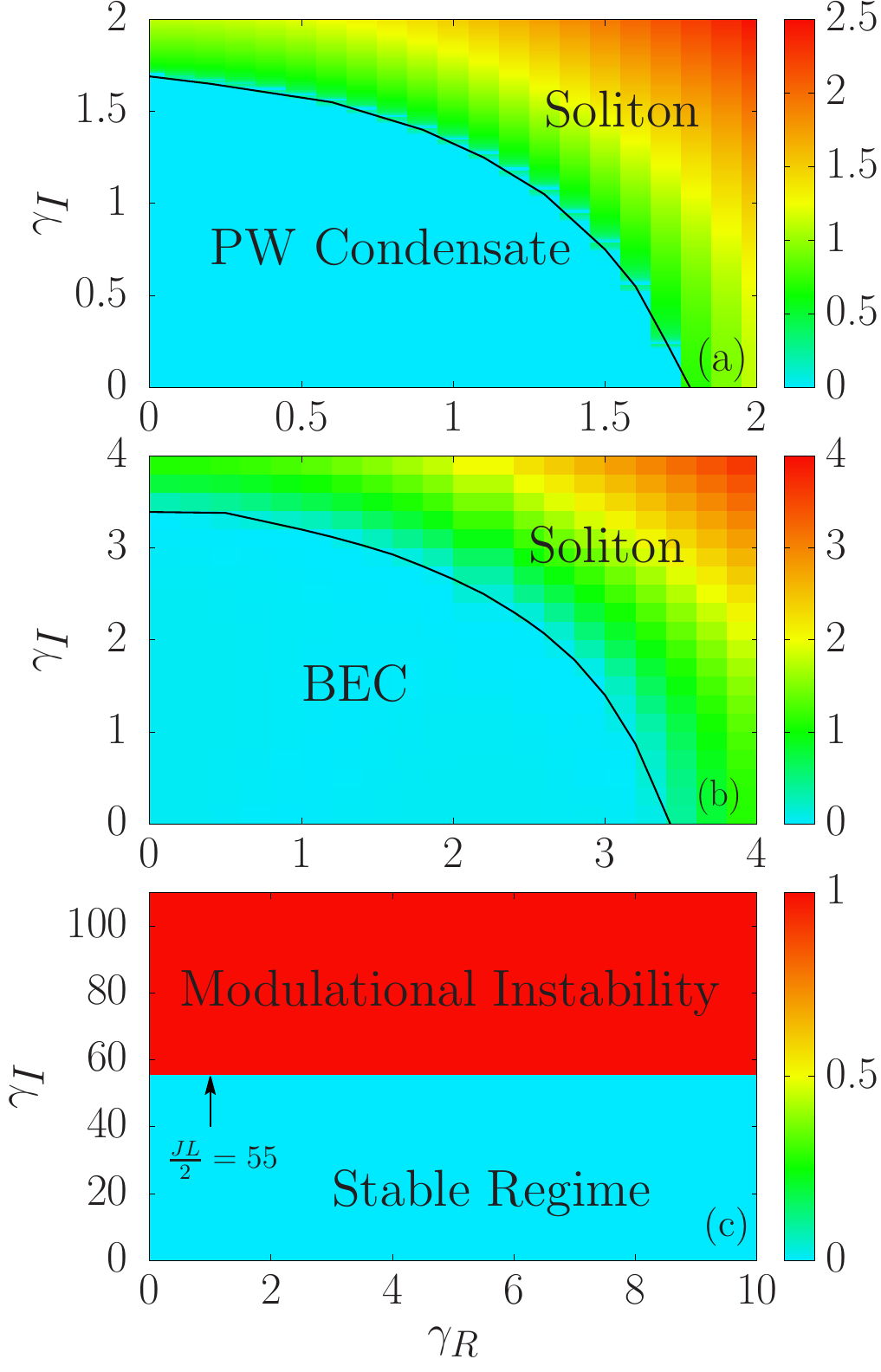}
    \caption{The phase diagram in the complex gauge coupling plane, based on the quasiparticle mode energy gap $\Delta E$ for (a) non-interacting and (b) interacting systems. The black phase boundaries separate PW condensate or BEC and localized soliton phase. The change of $\Delta E$ in $\gamma$-plane from a constant zero value to a finite gradient, distinguishes the two quantum phases. (c) shows the complex fraction of excitation energy in momentum space [Eq.~\eqref{eq.dispersion}]. The zero value corresponds to purely real energies (stable regime) while the purely imaginary $\epsilon(k)$ results in unity complex fraction which marks the modulational instability in the system. The instability occurs for $\gamma_{I} > JL/2 = 55$. Here, the system size $L=110$ and $J=1$.}  
    \label{fig:phase}
\end{figure}
 
\subsection{Dispersion curves}
The dispersion relation reveals the response of the system to external perturbations. For a condensate phase in the 1D optical lattice, the free-particle dispersion relation with respect to band minima is $\epsilon(k) = 2J(1 - \cos ka)$. To numerically obtain the dispersion curves of the quasiparticles, we perform the Fourier transform of the quasiparticle amplitudes to obtain $\tilde{u}^l_j(k)$ and $\tilde{v}^l_j(k)$ and compute the expectation value of the momentum of each quasiparticle. In momentum-space representations, the normalized root-mean-squared momentum weighted over the entire mode amplitude distribution is \cite{wilson_10,suthar_16}
\begin{equation}
   \langle k_j\rangle_{l} = \left[ \frac{\sum_{k_j} k_j^{2} \left( \left| \tilde{u}_{}^{l}(k_j) \right|^{2} + \left| \tilde{v}_{}^{l}(k_j) \right|^{2} \right) }{\sum_{k_j} \left( \left| \tilde{u}_{}^{l}(k_j) \right|^{2} + \left| \tilde{v}_{}^{l}(k_j) \right|^{2} \right) } \right]^{1/2}.
\end{equation}
The discrete Bogoliubov dispersion curves are then obtained by relating $\langle k_j\rangle_{l}$ to the excitation energy. The gauge field determines the dispersive properties of the model Hamiltonian. To examine the role of $\gamma$, we study the dispersion curves at two representative values of $\gamma$, corresponding to the condensate and soliton phases. For a fixed $\gamma_{R}=0$, at 
lower $\gamma_I$ corresponding to BEC phase, the low energy collective excitations exhibit a well defined quadratic behaviour of monotonically increasing energies at low $k$. The curve is a smoothly increasing dispersion relation typical of a stable, extended plane-wave condensate in an optical lattice. The curves are shown for $\gamma = 1.0i, 2.0i$, and $3.0i$, which  perfectly overlap, showing that the weak density-dependent gauge field does not alter the dispersive behaviour of the system.

Above a critical value of the imaginary gauge field strength ($\gamma_I > 3.39$), the ground state undergoes a first-order phase transition into a highly localized bright soliton. This radically alters the excitation spectrum. Unlike the BEC phase where low-momentum excitations cost near-zero energy, the entire spectrum shifts upward in $E_{l}$ as $\gamma_I$ increases from $4.0$ to $5.0$. This is due to the fact that the gauge field induces a strong reduction in effective inter-site tunneling, and pinning the soliton to a few lattice sites. Displacing or exciting a pinned, localized wavepacket requires a higher energy cost. The excitation energy gap increases with $\gamma_{I}$. These behaviour are consistent with the quasiparticle mode evolution shown in Fig.~\ref{fig:mode_U}. Furthermore, the lowest zero-energy mode acquires finite momentum in the soliton phase, as evident from the dispersion curve~\ref{dispersion_curves}(b). The finite value $\langle k \rangle$ moves to higher momentum for the increase in $\gamma_{I}$. The density-difference term lowers the kinetic energy if the soliton acquires a phase-gradient and carries a finite average momentum. This is analogous to the spin-orbit coupled system where the single-particle dispersion has minima at finite momentum~\cite{li_12,suthar_21}.  

\subsection{Phase diagrams}
 We further map-out the phases of the model in a phase diagram in the complex $\gamma$ plane. As discussed previously, the quasiparticle energies in the evolution of Bogoliubov modes depend on both components of the gauge field. The behaviour of observables such as chemical potentials and mode energies with complex $\gamma$ allows us to characterize quantum phases and the stability regime. The increase in one of the components of $\gamma$ requires the other component to be lower for the quantum phase transition. The ground-state phase diagram in the complex $\gamma$ plane is shown in Fig.~\ref{fig:phase}. To identify the phase boundaries, we compute the quasiparticle mode gap between the lowest and first excited mode $\Delta E$, both for non-interacting and interacting cases. In the upper $(U=0)$ and middle $(U=0.05)$ panels of the figure, the mode gap is presented, where a black line guides a phase boundary between the plane-wave condensate or BEC phase and a localized soliton phase. The soliton state is characterized by a finite gap while it remains (nearly) zero in the condensate phase. The gap in mode energies increases with $\gamma_{R}$ and $\gamma_{I}$. The critical gauge coupling increases with the onsite repulsion, as $U$ opposes the density localization required for the formation of the soliton. With $U$, the interaction energy of the localized state increases, and a stronger field is needed to compensate the energy cost and stabilize the soliton, thus shifting the BEC-to-soliton transition to a larger $\gamma$. [cf. Fig.~\ref{fig:phase}(a) and Fig.~\ref{fig:phase}(b)]. In particular, the transition occurs at $\gamma_R = 1.78$ with $\gamma_I=0$ or $\gamma_I =1.69$ with $\gamma_R=0$, suggesting the role of individual components in hosting the soliton state. With $U = 0.05$, the critical values shift to higher values at $\gamma_R = 3.39$ and $\gamma_I = 3.45$, respectively. The phase diagram of the non-interacting system agrees well with previous study~\cite{faugno_23}. Similar behaviour is shown by the change in $\mu$ in the complex $\gamma$-plane. 
 
 The dispersion relation of the Hamiltonian [Eq.\eqref{eq.dispersion}] allows us to parameterize the stable regime of the system. At $\gamma_{I} = JL/2$ the excitation energy in $k$-space becomes zero and above this critical value the excitation energy becomes complex. The complex quasiparticle energies or imaginary $\epsilon(k)$ are the signature of modulational instability in the system \cite{faugno_23}. To quantify the onset of modulation instability, we illustrate the complex fraction of the excitation energy $\epsilon(k)$ in the complex $\gamma$ plane in Fig.~\ref{fig:phase}(c). This fraction is the ratio of the imaginary part of the excitation energy and the energy magnitude. In the stable regime, $\epsilon(k)$ is purely real, which makes zero fraction. For $\gamma_{I} > JL/2$, $\epsilon(k)$ becomes purely imaginary, leading to a unity fraction. We mark the instability regime in the phase diagram [Fig.~\ref{fig:phase}(c)] at $\gamma_{I} = 55$ that remains independent of $\gamma_{R}$, confirming that the dispersion is affected only by the imaginary component $\gamma_{I}$ and size of the system $L$. 
  
\subsection{Real time dynamics}
\begin{figure}[h]
  \includegraphics[width=\linewidth]{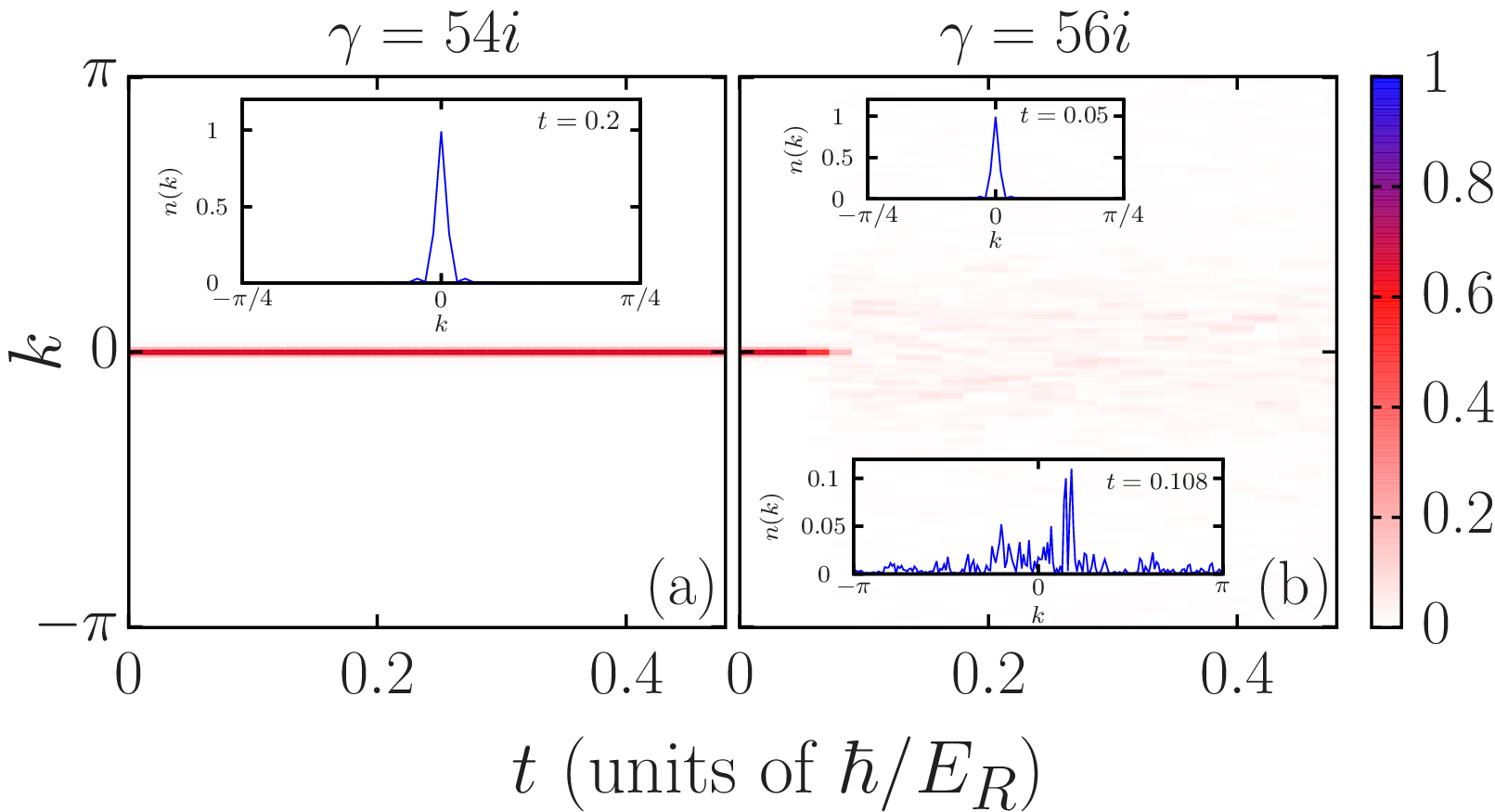}
  \caption{The time evolution of the momentum space correlation function for (a) $\gamma=54i$ and (b)=$56i$, representing gauge coupling below and above the threshold value of instability 
           $\gamma_I=JL/2$. In (a) the system shows a central peak at zero momentum $k=0$. In (b) the spread of correlation at $t\approx0.1$ signals the loss of coherence due to instability. The 
           insets show the momentum space distributions at selected times of the evolution.}
    \label{fig:mom_distri}
\end{figure}

Finally, we discuss the dynamical measure of stability and quench dynamics in the presence of a density-dependent gauge field. We first examine the unstable regime through the dynamical response of off-diagonal correlation at two representative values of $\gamma$. The  momentum space correlation is $\mathcal{C}(k) = L^{-1}\sum_{j,j'} e^{ik(z_{j} - z_{j'})} \Psi^{*}_{j}\Psi_{j'}$, where $z_{j}$ and $z_{j'}$ are two different lattice sites \cite{suthar_22}. The evolution of $\mathcal{C}(k)$ identifies the modulational instability in the system, as shown in Fig.~\ref{fig:mom_distri} for a non-interacting uniform system. 
We initialize the plane wave with random noise and monitor the free temporal evolution. For $\gamma_{I}\leqslant JL/2$, we find that the $\mathcal{C}(k)$ peak at zero momentum and the correlation remains steady at a longer time evolution. On the other hand, for $\gamma_{I}>JL/2$ the correlation spreads and vanishes at longer times, signaling nonlinear dephasing and loss of coherence. The dominant zero momentum peak distributed among various finite momentum modes in the unstable regime, as evident from the momentum distributions shown in the insets at selected time.
The spread of the $k$-space correlation is a consequence of instability. This is consistent with the static observable (dispersion relation) that predicts the instability in the $\gamma$-plane, presented in Fig.~\ref{fig:phase} (c).   

We further perform the dynamical quench for two $\gamma$ values corresponding to the condensate and soliton phases, for which the initial state is prepared using imaginary-time propagation. The onsite repulsion is set at $U=0.05$. In particular, we examine two dynamical phenomena, namely Bloch oscillations and breathing oscillations of a system in optical lattices.  
\begin{figure}[h]
  \includegraphics[width=\linewidth]{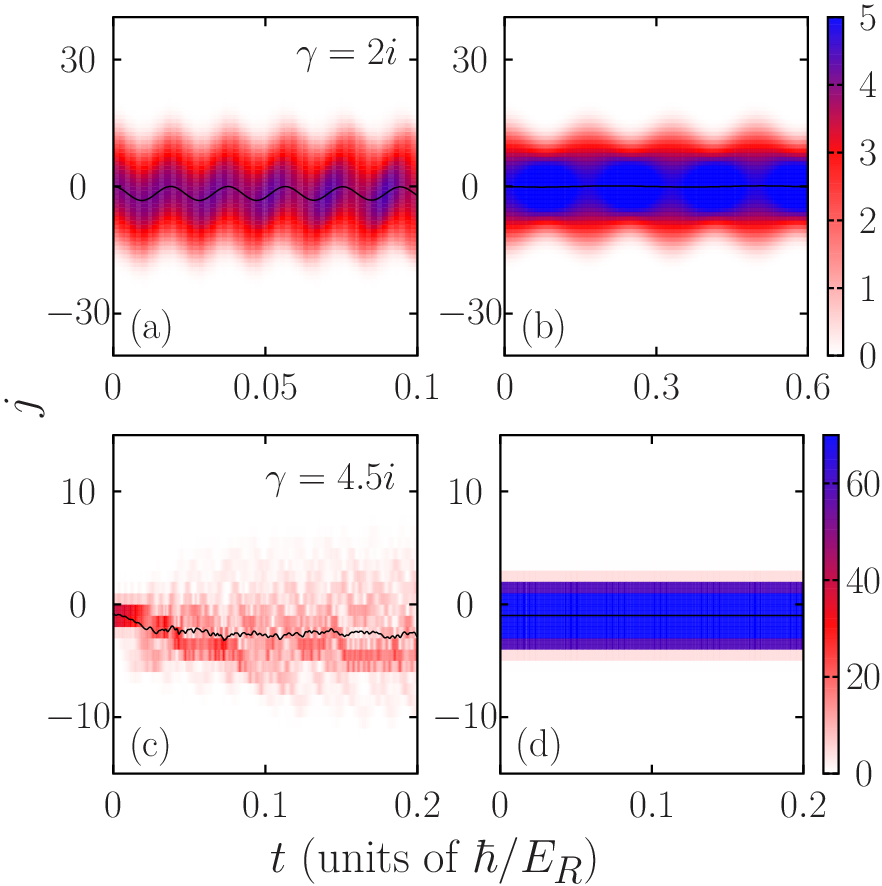}
  \caption{(a,c) The time evolution under perturbation of a constant linear external force and switch-off of a harmonic potential. (a) The BEC phase with $\gamma = 2i$ exhibits Bloch oscillations at a constant frequency while the condensate cloud maintains its Gaussian profile. (c) In the soliton phase with $\gamma = 4.5i$, the oscillations are destructed and the system does not return to initial position after one Bloch period. (b, d) The response of the system to a sudden quench in the harmonic trap strength. (b) The condensate exhibits breathing oscillations in optical lattice characterized by periodic compression and expansion cycles. (d) While the soliton phase displays complete dynamical rigidity, showing no observable structural changes over long time evolution. The solid line represents the center-of-mass position of the system in the dynamical evolutions.}
  \label{fig:real_time}
\end{figure}

{\textit{Bloch oscillations}}: The weakly-interacting ultracold atoms in optical lattices exhibit Bloch oscillations. To ensue such oscillations, we remove an external harmonic confinement and apply a linear external force. We first examine the dynamics of the BEC phase for $\gamma=2i$. The dynamical evolution of the condensate density is presented in Fig.\ref{fig:real_time}(a), which
shows the Bloch oscillations with a constant frequency and preserves the Gaussian character in the absence of an external trap. This is also evident by the expectation of the position that guides the oscillations of the condensate. Fig.\ref{fig:real_time}(c) displays the dynamics corresponding to a soliton phase corresponding to $\gamma=4.5i$, which shows the destruction of oscillations. 
Due to the presence of DDT, the soliton does not behave like a rigid wave-packet. As a result, the translational motion becomes coupled with the internal degree of freedom of the soliton, leading to asymmetric motion and the eventual loss of periodic oscillations. 

{\textit{Breathing oscillations}}: We further introduce breathing oscillations by applying a sudden perturbation to the external harmonic trap. The suppression of the expansion of the density of the bright soliton of attractive bosons in a 1D optical lattice above a critical interaction strength has previously been reported~\cite{naldesi_19}. The expansion dynamics with an increase in trap strength by $50\%$ is shown in Fig.\ref{fig:real_time}(b) and Fig.\ref{fig:real_time}(d) for the BEC and soliton phase, respectively. The density in the condensate phase exhibits periodic compression and expansion of clouds. In contrast, the density remains completely steady over time in soliton phase. This density invariance implies that the soliton is pinned to its position, demonstrating a negligible role of an external trap quench in its structure. This is consistent with the large quasiparticle energy gap in the soliton phase, which inhibits the breathing oscillations or does not affect the atomic density.  

\section{Conclusions}\label{conclusion}
We have studied the low-lying quasiparticle excitations of ultracold bosons with density-dependent hopping in a one-dimensional optical lattices. The evolution of quasiparticle energies obtained using the Bogoliubov theory of D$^4$NLSE exhibits a mode gap across the first-order quantum phase transition between the Bose-Einstein condensate and the bright soliton. The finite mode gap widens with gauge coupling of the density-difference term as the density imbalance enhances the localization and suppresses the low-energy excitations. The onsite repulsion shifts the gauge coupling of the condensate-to-soliton phase transition to a larger value because it introduces interaction-induced energy cost for the density fluctuations and thereby increases the effective restoring force against excitations. We have further mapped out the parameter space of quantum phases and the stability regime based on the quasiparticle mode gap and dispersion relation. The energy minimization requires the zero-energy mode of the soliton to acquire a finite momentum. Finally, the quench dynamics delineate the suppression of Bloch oscillations and absence of breathing oscillations for a stronger gauge field of the soliton state. The recent progress on ultracold experiments of the synthetic gauge field and Floquet engineering provides a promising pathway to realize the density-driven novel quantum phenomena and quasiparticle excitations.

\begin{acknowledgements}
We thank H. Kaur and N. Kaloya for the valuable discussions. P.K.M. acknowledges financial support from the University Grant Commission (UGC), New Delhi. K.S. acknowledges support from the Science and Engineering Research Board, Department of Science and Technology, Government of India through Project No. SRG/2023/001569. 
\end{acknowledgements}

\bibliographystyle{apsrev4-2} 
\bibliography{references}
\end{document}